\documentclass[10pt,a4paper]{article}
\title{NMR AND THE ANTIFERROMAGNETIC CRYSTAL PHASE REGIONS IN RAPIDLY QUENCHED RIBBONS AND IN ALLOYS OF THE TYPE $Cu-Mn-Al$.}
\author{Matej Hudak \\ {\it Lab,} Stierova 23, SK - 040 23  Kosice, Slovak Republic  \\hudakm@mail.pvt.sk
\and Jana Tothova\\ {\it Lab,} Stierova 23, SK - 040 23  Kosice, Slovak Republic
\and Ondrej Hudak \\  {\it Technical University of Kosice},\\ Faculty of Aerodynamics, Department of Aviation Technical Studies,\\ Rampova 7, SK - 040 01  Kosice, Slovak Republic\footnote{Corresponding author}\\
{hudako@mail.pvt.sk}}
\date{April 11, 2018}
\begin{document}
\maketitle{}

\section*{PACS Numbers:}
\begin{itemize}
\item 
75.20.En 	Metals and alloys
\item 
76.50.+g 	Ferromagnetic, antiferromagnetic, and ferrimagnetic resonances; spin-wave resonance
\item 
76.60.-k 	Nuclear magnetic resonance and relaxation
\end{itemize}
\section*{Condensed paper title:}
{\it NMR AND AFM REGIONS IN RAPIDLY QUENCHED RIBBONS AND ALLOYS $Cu-Mn-Al$.}
\newpage
\vspace{1in}
\begin{abstract}
It was shown that anomalous resistivity behavior of the $Cu-Mn-Al$ ribbons is explained by the s-d interaction between conduction electrons and the clustered Mn atoms. While nuclear magnetic resonance measurements show the antiferromagnetic and ferromagnetic clusters of Mn atom coexisting without long-range order, it is an interesting problem to study magnetic resonance properties also for the antiferromagnetic crystal phase regions (which have long-range order for larger regions) and which may also occur in these ribbons. The Heusler  Type  $Cu-Mn-Al$  Alloy has  a  composition  half way  between $Cu_{2}MnAl$ and $Cu_{3}Al$. Electron  microscopy  of  the  premartensitic  $\beta Cu-Zn-Al$  alloy has shown that the  $\beta Cu-Zn-Al$  alloy  quenched  from  high  temperature  has the  electron  diffraction  patterns  of  this  alloy  well  explained  by  the model  with   the  existence of  small  particles  with  an  orthorhombic  structure.  It was noted that an important aspect of improvement in the material properties is to create a nanostructured state in matrix, which has significant advantages in magnetic and mechanical characteristics in contrast to the bulk materials in crystalline or amorphous state. It is an interesting problem to study magnetic resonance properties not only for the Mn atoms and clusters without long-range order but also for the antiferromagnetic crystal phase regions (which have long-range order for larger regions) which may also occur in ribbons. This is the aim of our paper.
\end{abstract}
\newpage

\tableofcontents{}

\newpage
\section{INTRODUCTION.}

Metallic glasses \cite{CS} are a class of metallic materials that do not display long-range atomic order. Their amorphous character and the lack of dislocations, these materials exhibit mechanical properties that are quite different from those of other solid materials \cite{1.2} and \cite{2.2}. They have also interesting physical and chemical properties. Some metallic glasses exhibit superior soft magnetic properties \cite {3.2}, good magnetocaloric effects \cite{4.2}  and outstanding catalytic performance \cite{5.2}, thus having potential for a widespread range of technological applications \cite{6.2}. Metallic glasses when heated in the supercooled liquid region allow moulding and shaping with microscale precision by means of thermoplastic processing [7] \cite{7.2}. This has lead to the development of diverse products based on these alloys (sporting goods, medical and electronic devices and advanced aerospace applications). In spite of their large elasticity,  metallic  glasses  exhibit  poor  room-temperature  macroscopic  plasticity  compared  to polycrystalline metals \cite{8.2}. This low plastic deformation, particularly evidenced when testing metallic glasses under tension, is related to the formation and rapid propagation of shear bands \cite{2.2}.

Therefore \cite{CS} routes  to  enhance  plasticity  of  metallic  glasses  include  procedures  to  hinder  shear  band propagation. This can be achieved, by designing composite materials consisting of particles which act as second-phase reinforcements embedded in the amorphous matrix \cite {9.2} and \cite{10.2}. Other approaches 
towards toughening of metallic glasses have also been developed, such as the preparation of the so-called dual-phase amorphous metals \cite{11.2}, some specific surface treatments (e.g., laser or shot pinning) \cite{12.2}.

In his PhD. work V. Ocelik, \cite{OHD}, supervised by one of the authors (O.H.) formation and rapid propagation of phase change from amorphous to crystalline was studied, and described were nonhomogeneous plastic deformation and defects in metallic glasses. These defects contribute to the macroscopic rapid propagation of shear bands. Ocelik (et al.), also in \cite{NOPH} studied influence of laser treatment. Based  on  a  recently  published  recursive  model  describing  the  geometry  of  laser  clad  coatings  and  on  experimental  track characteristics the authors propose specific functions to describe the geometry of laser clad coatings formed by overlap of individual tracks depending on the processing parameters. This work was extended in \cite{OBDH}, where the failure of amorphous metallic materials under uniaxial tensile stress at temperatures much lower than the glass transition temperature was studied. It is known that it is preceded by an intense shear deformation localized into narrow bands. These bands lie near the planes of maximum shear stresses. Significant changes in the original structure, taking place during the process of a local shear deformation inside the bands, leads to a considerable decrease in the viscosity. Because of this, the sample will fail under the influence of tensile stress in one of these shear bands. The crack propagates alternatively in two equal-stress planes of maximum shear stresses, which are perpendicular to each other. In \cite{OBDH}, where there were done statistical investigations of fracture demonstrations on Ni-Si-B metallic glass ribbons failed in tension at 4.2 to 300K, two alternative quantitative mathematical descriptions of the yield stress anisotropy in the plane of amorphous alloys ribbons are proposed in this paper, that are based on two models: model of the plane stress state and model of the oriented anisotropic polyatomic clusters. These descriptions give adequate approximations of the experimental angular dependences of the yield stress for some amorphous alloys. This models were used to explain phenomena in ribbons of the type  $Ni_{80}Si_{10}B_{10}$ and $Ni_{80}Si_{5}B_{15}$ for determination of temperature dependence of propagation of defects.

In \cite{HCMJDO} the  metallic  ribbons  $Fe_{40}Ni_{40}B_{20}$ and  $
Cu_{47}Ti_{35}Zr_{11}Ni_{6}Si_{1}$  and  $Zr_{65}Cu_{17.5}Ni_{10}Al_{7.5}$   with  different microhardness and glass forming ability were studied at different loading rates from 0.05 to 100 mN/s. Authors describe in details the differences in elemental discontinuities on the loading curves for the studied alloys. They have found that the discontinuities began at a certain local deformation independently on the macroscopic mechanical properties of a ribbon.

A review how the presence of interstitially defects and the change in their concentration can induce structural relaxation when metallic glasses are heat treated to temperatures below or around the glass transition is in \cite{15.2}. Structural relaxation is important, it can induce changes in many physical  properties  of  the  glasses,  particularly  in  the  mechanical  behavior  (elasticity,  anelasticity,  
viscoelasticity,  etc. \cite{16.2}),  but  also  in  the  electrical,  corrosion  and  even  magnetic  performances.

In \cite{HGO} and \cite{GHHSHO} $Al $ nuclear-quadrupole-resonance studies of $CeAl_{2}$ in the temperature range between $0.09$ and $4.2 K$ were done.
We started from \cite{M} - \cite{Ca}.
 Below $T_{N}= 3.45 K$ (the Neél temperature)  $CeAl_{2}$ orders antiferromagnetically. The spin-lattice relaxation rate $\frac{1}{T_{1}}$ shows a sharp peak at $T_{N}$, and the NQR linewidth increases from $27±3 kHz$ in the paramagnetic state to $270±15 kHz$ at $2 K$. Below $1.5 K$, in the low-temperature regime of the magnetically ordered state, the temperature dependence of the spin-lattice relaxation rate follows a Korringa law with $\frac{1}{(T_{1} T} = 7.9±0.8 \frac{1}{K sec}$. This relaxation rate is much larger than those found in normal metals. Above 1.5 K deviations from the Korringa law are observed. These deviations are consistent with a relaxation produced by magnonlike excitations with an energy gap of $11±3 K$, in agreement with results of previous neutron scattering and $T_{1}$ measurements. The NQR is here a method which enables to study properties of antiferromagnetic phase and a transition  to it from the paramagnetic phase.	The NMR method is usually used with measuring other properties of the material. For example thermodynamics of $CeAl_{2}$ at low temperatures was studied measuring specific heat and spin-lattice relaxation rate \cite{HGO}. Thermodynamic calculations show that a considerable fraction of the observed low-temperature nuclear spin-lattice relaxation rate and specific heat of $CeAl_{2}$-both varying linearly with temperature-may be due to the existence of new low-frequency features in the spin-excitation spectrum of incommensurate magnetic structures.
In \cite{HH} we studied magnetic nanoparticles with core shell structure. We formulated the macroscopic model and calculated the coercive field in these nanoparticles with nonhomogeneous structure. They were mechanosynthesized and consisting of an ordered core surrounded by the shell. The shell may be structurally and magnetically disordered, or it may be ordered. These nanoparticles are found to be roughly spherical. We formulate the macroscopic model for the description of magnetic properties of nanoparticles with core-shell structure. The case of spheroids oriented in the same direction of polar axes is considered. There exits two coercive fields. Thus we see that for above mentioned nanoparticles it is not possible to use the NMR theory as for crystalline particles, which are from the type of microparticles to larger particles. We will use in our paper the description of these later particles as those which may describe small crystals in ribbons and alloys.

Let us also note \cite{OH} that an induced magnetic anisotropy in amorphous ribbons may occure, we studied the case of negative magnetoelastic constant. The case of positive constant was studied elsewhere, see \cite{OH}.
One of the most important mechanisms for the appearance of the induced magnetic anisotropy in amorphous ferromagnetic materials is due to internal stresses in the case of nonzero magnetoelastic coupling. It has,been proposed that according to different ways of solidification in diferent regions, the amorphous ribbon may have four types of these solidification regions. The distribution of internal stresses may be specified for each of these regions. Then the orientations of the magnetization of the ribbon may be specified from the magnetoelastic energy. We have found that 
\cite{OH1} for amorphous materials the spins are in disordered state also in nonzero magnetic field, besides some isolated spins. Local internal fields in vector spin glasses with random uniaxial local anisotropy are shown to be positive on all non-isolated spins whenever the spin state corresponds to any minimum of configurational energy within the classical theory. Moreover it is shown that there exists a lower positive bound on the values of the local internal fields. This bound depends on the exchange interaction constants and on the strength of the local anisotropy. Thus in amorphous ribbons with regions of crystalline phase the AFM state should be studied using theory of NMR for the AFM phase to distinguish it from the disordered regions of amorphous ribbons.

 However  as second-phase reinforcements embedded in the amorphous matrix may be present  antiferromagnetic phase, which may be hard to detect by macroscopic magnetic measurements. Multicomponent bulk metallic glasses (BMG) have attracted great attention \cite{HSOMA} because  of  their  unusual  physical,  chemical  and  mechanical  properties  \cite{1}. Mechanical relaxation of metallic glasses was overviewed (experimental data and theoretical models) in \cite{LPC}. 
 In the paper \cite{KKK} the authors refer about anomalous electrical resistivity and nuclear magnetic resonance in rapidly quenched Cu-Mn-Al ribbons. In the $Cu_{32}Mn_{35}Al_{33}$ ribbon, it is found \cite{KKK} by nuclear magnetic resonance measurements that the antiferromagnetic and ferromagnetic clusters of Mn atom coexist without long-range order. Authors show that anomalous resistivity behavior of the $Cu-Mn-Al$ ribbons is explained by the s-d interaction between conduction electrons and the clustered Mn atoms. While \cite{KKK} found by nuclear magnetic resonance measurements the antiferromagnetic and ferromagnetic clusters of Mn atom coexist without long-range order it is an interesting problem to study magnetic resonance properties not only for the Mn atoms and clusters without long-range order but also for the antiferromagnetic crystal phase regions (which have long-range order for larger regions) which may also occur in these ribbons. This is the aim of our paper.

\section{RAPIDLY QUENCHED Cu-Mn-Al RIBBONS, AND ALLOYS.}

In the paper \cite{KKK} the authors refer about anomalous electrical resistivity and nuclear magnetic resonance in rapidly quenched Cu-Mn-Al ribbons. As they write the rapidly quenched $Cu_{67-x}Mn_{x}Al_{33}$(at.percent) ribbons show high electrical resistivities and low temperature coefficients of resistivity. The resistivities
decrease monotonically with increasing temperature for the ribbons of more than $30 $(at.percent) Mn concentration. In the $Cu_{32}Mn_{35}Al_{33}$ ribbon, the author found by nuclear magnetic resonance measurements that the antiferromagnetic and ferromagnetic clusters of Mn atom coexist without long-range order. The anomalous resistivity behaviour of the Cu-Mn-Al ribbons is explained by the s-d interaction between conduction electrons and the clustered Mn atoms. Let us note that in \cite{BLT} for the $Cu-Mn-Al$ detailed  electron metallographic  studies  were made  of interfacial  dislocations  which  are  formed  to  relieve  the  elastic  coherency strains  developed  upon  long  aging  of  spinodal  alloys  in the  system.  Interfacial dislocations  in the  Heusler  type alloy  $Cu-Mn-Al$  which  also  appears  to  undergo  spinodal  decomposition \cite{7}, and in recent paper \cite{7.1}. As authors \cite{BLT} have found  for the Heusler  Type  $Cu-Mn-Al$  Alloy the  alloy  studied has  a  composition  half way  between $Cu_{2}MnAl$ and $Cu_{3}Al$.  Upon  quench-aging  possesses  all the  metallographic  characteristics  of a  spinodal  decomposition. The  ternary constituent has the $L2_{1}$ structure  and  the  binary  constituent has  the  closely  related $D0_{3}$. Their  lattice parameters  differ by $2$ percent. The $Cu-Mn-Al$  alloy developes  interface  dislocations  lying  in pure edge  orientation  in the ${001}$  interface  planes  and  with $( 100 )$  Burgers vectors. In rapidly quenched Cu-Mn-Al ribbons with larger regions of antiferromagnetic or ferromagnetic phase the interface may exists with interfacial defects (Ocelik´s study of defects in amorphous ribbons). This is due to the fact that a restriction in the  $Cu-Mn-Al$  spinodal  alloys  is that both  phases  are  ordered. In ribbons amorphous phase and ordered phase may  exist. 
As noted by \cite{TDPG} and \cite{3.1} - \cite{3.3} in the market of new materials, the functional materials having unusual properties are in great demand, among which the ferromagnetic shape memory alloys are predominant. The control over such properties is exercised using force, thermal, and magnetic fields. As electron  microscopy  of  the  premartensitic $\beta  Cu-Zn-Al$  alloy has shown that \cite{MDSD} the  $\beta Cu-Zn-Al$  alloy  quenched  from  high  temperature  studied by  electron  microscopy has the  electron  diffraction  patterns  of  this  alloy  well  explained  by  the model  proposed  by  \cite{TS},  i.e.  by  the  existence of  small  particles  with  an  orthorhombic  structure, see also \cite{3.21}.  This  structure  is,  however,  closely  related  to  the ω-structure. Thus it was noted in \cite{TDPG} an important aspect of improvement in the material properties is to create a nanostructured state, which has significant advantages in magnetic and mechanical characteristics in contrast to the bulk materials in crystalline or amorphous state. Magnetization and magnetic anisotropy in case of nanoparticles can be significantly greater than that of a bulk sample, and a difference between the Curie temperature $( T_{c} )$ reaches hundreds of degrees \cite{3.1} - \cite{3.3}.

To study clusters, and even possibility of existence of regions (nanocrystals) of crystal phase in amorphous ribbons or in a matrix specified by another phase, different measurements of magnetic properties of are used. Measurements of magnetostriction are described in \cite{GSHP}. A  survey  of  different  measuring  methods  suited  for  soft  magnetic  materials  is  given. The methods are subdivided into direct and indirect methods. The SAMR method (Small angle magnetization rotation) is best suited especially for low magnetostrictive ribbons. The author discusses using these methods on  example  selected  results  and  an analysis of the temperature dependence of the magnetostriction as measured on amorphous $Fe_{85-x}Co_{x}B_{15} $. 
Magnetic nanomaterials have a number of unusual properties,in particular, giant magnetoresistance, abnormally large magnetocaloric effect, and others \cite{3.4}. 
$Cu – Al – Mn$ alloys are one of the most interesting ferromagnetics with shape memory (SM). They demonstrate an unusual magnetic behavior in superparamagnetism \cite{3.5} and giant magnetoresistance \cite{3.6}, and specific mechanical properties such as SM effect, thermoelasticity, superelasticity, and plasticity of transformation \cite{3.7} and \cite{3.8}. They exhibit a superelastic strain of about $7$ percent , which is comparable to that of $Ti – Ni$ alloys \cite{3.9} and \cite{3.10}. To get optimal properties, these alloys undergo an additional thermal, mechanical, or magnetic treatment. Aging of $Cu – Al – Mn$ alloys leads to the formation of a system of nanoscale particles of ferromagnetic $Cu_{2}MnAl$ phase in a paramagnetic $Cu_{3}Al$ matrix \cite{3.5}, and annealing in magnetic field increases the $T_{c}$ of $Cu – Al – Mn$ alloys \cite{3.14}. At the same time, the heat treatment allows to control number and size of particles in the alloy and also the martensitic transformation temperature and hysteresis, which depend on characteristics of precipitated particles \cite{3.15} and \cite{3.16}. The clarification of the possibility to control the magnetic and mechanical characteristics of $Cu – Al – Mn$ alloys and amorphous ribbons (under annealing in zero- or non-zero magnetic field) is of interest.

\section{NUCLEAR RESONANCE IN ANTIFERROMAGNETS - MOTIVATION.}

Nuclear resonance as is described in the Jaccarino's paper \cite{J}  
has its criteria for its observing nuclear resonance in magnetic solids. They are connected with the nuclear hamiltonian. It is described in general for nonmagnetic solid and  magnetic solid via study of internal fields.

In antiferromagnets we have to identify the spin lattice and its symmetry.
We will describe macroscopic and microscopic properties by some remarks.
It is important to find ground states, excitations, describe phase transitions which may occur. For nuclear relaxation the relaxation and its linewidth is important.
Firstly we will describe the relaxation and its linewidth  in paramagnetic phase, and then we will study local fields in AFM ( lineshape, spin-relaxation rate ), with 	indirect nuclear spin-spin interaction. A possibility of incommensurate antiferromagnetic structure is taken into account.
Why are AFM phase regions interesting nowadays for the rapidy quenched ribbons: see above \cite{KKK}. However in theory the nature of the ground state in quantum AFM \( d>1 \) is not understood well. Besides exotic phases ( spin liquids, spin nematics, ...)
antiferromagnetic periodic and quasiperiodic systems may occur, namely also in mettalic glasses. In fact first	experiments with nuclear resonance in antiferromagnetic phase observed for water protons of	\( CuCl_{2}.2H_{2}O \) at \( T_{LHe} \) were done by \cite{PH}. However sometimes
nature of the ground state in quantum AFM is not understood well for ( \( CeAl_{2} \) \cite{HGO} and \cite{GHHSHO} and other heavy fermion systems, high-temperature superconductors, 	quasicrystals, ... . Why NMR measurements are interesting for study of rapidly quenched ribbons? It is well known plus this is weakly interacting probe of internal fields.
And why NR in AFM?	Static (time-averaged magnetic field) and dynamic properties of ordered phase ( some fluctuation components) may be studied.

\section{CRITERIA FOR OBSERVING NUCLEAR RESONANCE IN MAGNETIC SOLIDS.}

Magnetic solids we call those solids in which magnetically ordered 
phase (FM, AFM, SDW, ...) occurs. Electronic dipolar fields in mg. solids are \( \approx 10^{3} \) times larger than corresponding nuclear dipolar fields, atomic hyperfine fields of magnetic ions are \( \approx 10^{6} \) times larger. Does it mean that  resonance frequency is 	\( \approx 10^{6} \) larger ? No: the integrated spectral density is \( \approx \) constant + distribution of the local field spectra due to exchange over large frequency range \( \approx \) the \( \mid H_{eff} \mid \ll
	H_{static} \rightarrow \).
Crude necessary condition for observing NR in a magnetic solid for \( T \geq T_{N} \) is:
\[ \frac{1}{T_{1,2-min}} \geq \frac{ ( \gamma H_{int} )^{2}}{ \omega_{e}}, \]
where \( T_{1,2-min} \) is the smallest value for which NR is still observable, here
\( \omega_{e} \) is the exchange frequency, \( H_{int} \) is the static value of the perturbing field at the nucleus.

In the ordered phase \( T \leq T_{N} \) there is more complex criterium which
holds for \( T_{2} \). Difference from the paramagnetic case is that in AFM the electron spins do not reorient themselves rapidly, thus  the constant in time field \( \approx \) the instantaneous value of the electron	field.
Difference from the paramagnetic case is that two nuclear spins which have 
equivalent positions in the crystal do not have necessarily the same magnetic characteristics in AFM.

\section{NUCLEAR HAMILTONIAN.}

In nonmagnetic solid: \[ {\it H} = \gamma \hbar {\bf I.H}_{0}, \]
where {\bf I} is the nuclear moment, \( {\bf H}_{0} \) is the external field, there are 
small corrections due to dipolar fields of other nuclei, atomic diamagnetism, 
and chemical shift. In magnetic solids electronic spins rapidly fluctuate due to
exchange, dipolar interactions, spin-lattice interactions and spin-quasiparticle interactions. The nuclei in a static field \( \approx < {\bf S} > \) are related to the uniform	magnetization per unit volume \( {\bf M}_{0} \):
	\[	{\bf M}_{0} = Ng \beta < {\bf S} >_{T,H,p,...},   \]
where N is the volume density of spins, g is the electronic gyromagnetic ratio (tensor),
\( \beta \) is the Bohr magneton.
Internal fields of nuclei of nonmagnetic atoms are  ( \( H^{1} \) in \( CuCl_{2}.2H_{2}O \) ):
\[	{\it H}_{k} = - \gamma \hbar {\bf I}_{k}.( {\bf H}_{0} -
	g \beta \sum_{n} r_{n}^{-3} ( <{\bf S}_{n}> - \frac{3 {\bf r}_{n}( {\bf r}_{n}.
<{\bf S}_{n}>)}{r_{n}^{2}})). \]
Here we have (2-nd term) the dipolar field \( H_{D} \), which is in general not parallel to $\bf H_{0}$ for \( n \not= k \). Note that 
crystalographicaly inequivalent sites have different dipolar fields (low symmetry crystals, 2 and more inequivalent sites/unit cell ), so there is a number of different transitions \( \omega_{i} ( \angle {\bf H}_{0} ({\bf a, b, c}) ) \), here
g is a tensor with am orientation, temperature and field dependence.

For nuclei of magnetic atoms ( \( Co^{59} \) in \( CoF_{2} \) ) we have:
\[	{\it H}_{k} = - \gamma \hbar {\bf I}_{k}.( {\bf H}_{0} -
g \beta \sum_{n} r_{n}^{-3} ( <{\bf S}_{n}> - \frac{3 {\bf r}_{n}( {\bf r}_{n}.
<{\bf S}_{n}>)}{r_{n}^{2}})) + {\bf I}_{k}.{\it A}.< {\bf S}_{k}>, \]
where {\it A} is the electron-nuclear hyperfine interaction due to spin and
orbital moments of electrons of a given (paramagnetic) atom, and
\( {\bf H}_{HF}	= - \frac{ {\it A}.{\bf <S>} }{ \gamma \hbar } \), where
resonance frequency is:
\[	\omega = \gamma ( \sum_{i}^{x,y,z} (H_{0}^{i} + H_{D}^{i} +
H_{HF}^{i} )^{2} )^{ \frac{1}{2} } \]
and where g and {\it A} tensor axis coincide. Here:
\[	\omega( I_{z} \leftrightarrow I_{z}-1 ) = \frac{1}{ \hbar } {\it A}_{z}.
<S_{z}> - \gamma ( H_{D} \pm H_{0} ) + \frac{ 3 e^{2} q Q ( 2 I_{z} - 1)  }{4
\hbar I (2I - 1) }. \]
For nuclei of partially magnetic atoms ( \( F^{19} \) in \( MnF_{2} \) ) there is 
overlap between the wave functions of electrons of nominally nonmagnetic
ions and those of electrons of the paramagnetic ions, from this it follows 
redistribution of the spin magnetization.
An example is \( MnF_{2} \) where:
\[     < (2s)_{F^{-}}, up \mid (3d)_{Mn^{2+}}, down > =0,  \]
\[     < (2s)_{F^{-}}, up \mid (3d)_{Mn^{2+}}, up > \not= 0,  \]
orthogonalization leads to net 2s spin. Transfer of electrons of a given spin orientation from the spin-paired orbitals at the \( F^{-} \) to unpaired 3d orbitals on the \( Mn^{2+} \) leads to the nuclear Hamiltonian: 
\[	{\it H}_{k} = - \gamma \hbar {\bf I}_{k}.( {\bf H}_{0} + {\bf H}_{kD} )
+ \sum_{n}^{(nn)} {\bf I}_{k}.{\it A}_{n}.< {\bf S}_{n}>, \]
here we have transferred hyperfine interactions (last sum) and
resonance frequencies are:
\[	\omega_{ \pm } = \frac{1}{ \hbar } (2A^{z}_{I} - A^{z}_{II})
<S_{z}> - \gamma ( H^{z}_{D} \pm H_{0} ). \]
Numerical values of internal fields magnitudes of three kinds of internal fields are:
\begin{itemize}
\item	nonmagnetic ions ........... (dipolar fields \( 0.5 - 7.10^{3} Oe \) 
\item	magnetic ions ...............(mg. hfs fields \( 0.5 - 7.10^{5} Oe \) 
\item	partially magnetic ions .....(transfer hfs f \( 0.5 - 7.10^{4} Oe \) 
\end{itemize}

\section{ANTIFEROMAGNETS.}

We have a spin lattice in a crystal lattice in which the relevant degrees of freedom are individual atomic spins:
\[	{\bf S}_{j}..........{\bf R}_{j},     \]
where:
\[	[ S_{j \alpha} , S_{l \beta } ] = \delta_{jl} i \hbar \epsilon_{
\alpha \beta \gamma } S_{j \gamma }. \]
From symmetry considerations for the crystal we can determine structural symmetry 
where we take into account crystal and  spins, e.i. we use the magnetic symmetry. 
Macroscopic description of this system is based on the free energy F:
\[	F( {\bf M}, \epsilon_{ \alpha \beta },..., {\bf H}, \sigma_{ \alpha
\beta } , T ),   \]
where order parameter is magnetisation {\bf M}({\bf Q}), here $Q$ is wave vector,
there exists a coupling to other degrees of freedom in the crystal. We are looking for the free energy minimum which gives a ground state, then we study small oscillations of order parameters and from these spin waves, etc.

Microscopic description has similar principles. Heisenberg Hamiltonian is the simplest hamiltonian introduced by Dirac (1929):
\[	{\it H} = \sum_{ <i,j> } J_{ij} {\bf S_{i}.S_{j} } ,  \]
where: 
\[	J_{ij} \equiv J( \mid {\bf R_{i} - R_{j} } \mid ) . \]
Then we study CEF (crystal electric field) from the Coulomb interactions between each electron and all the charges lying around the ion, we obtain the electrostatic potential which gives splitting of energy levels. Also other interactions like
dipole-dipole interaction, are taken into account.

There are several ground states: the Neel state (simple, complicated) (f.e. in superconductors(AB), in $CeAl_{2}$, in a commensurate state: 
\[	{\bf M} = {\bf M}_{0} cos( {\bf k.r} + \phi )  \]
with the modulation wavevector {\bf k}. This later phase may be 
1, single-k structure 2,double-k structure 3, triple-k structure.
The condition for commensurability (sc lattice with  a - lattice const. ) is:
\[	{\bf k} = \frac{2 \pi }{a} ( \frac{M_{1}}{N_{1}}, \frac{M_{2}}{N_{2}},
\frac{M_{3}}{N_{3}} ),  \]
there exists at least one {\bf R} such that:
\[	{\bf k.R} = 2 \pi x INTEGER,  \]
and where M's and N's are integers, minimum of \( \mid R \mid \) gives new periodicity.
Polarization of this structure may be longitudinal \( {\bf M} \parallel {\bf k} \) or
transversal \( {\bf M} \perp {\bf k} \). 
This structure may be determined by diffraction from  Bragg peaks, it is 
pinned to the lattice and has an energy gap.

The incommensurate state is characterized by a modulation wavevector {\bf k}
which may be again 1, single-k structure 2, double-k structure and 3, triple-k structure.
Conditions of incommensurability of the structure (for sc lattice with a - lattice const. ) are:
\[	{\bf k} \not= \frac{2 \pi }{a} ( \frac{M_{1}}{N_{1}}, \frac{M_{2}}{N_{2}},
\frac{M_{3}}{N_{3}} ),  \]
there is {\bf no R} such that:
\[	{\bf k.R} = 2 \pi x INTEGER,  \]
where M's and N's are integers. The translational symmetry is lost, however
the polarization exists and is longitudinal or transversal, diffraction gives Bragg peaks which still exist, and the structure is not pinned to the lattice, there is continuous degeneracy of the ground state with  \( \phi \) arbitrary. 
There may exist domain walls: the ground state degeneracy (discrete, continuous ) lead to
the configuration of spins connecting different ground states, this is called a domain wall. We speak about solitons for some 1d walls. Note that the spin orientation changes in domain walls.

Antiferromagnetic excitations were studied in \cite{M}, \cite{W}, \cite{Akh}, \cite{Ke}, and \cite{Ca}.
Let us first discuss linear spin waves (LSW). We have a  bipartite lattice:
\[	A ... {\bf S}_{A}^{+} \mid 0> = 0, \]
\[	B ... {\bf S}_{B}^{-} \mid 0> = 0. \]
Diagonalization of the Hamiltonian ( Heisenberg with anisotropy):
\[	{\it H} = E_{GS} + \sum_{ {\bf k} } \epsilon ( {\bf k} ).
( a^{*}_{ {\bf k}} a_{ {\bf k}} + ( b^{*}_{ {\bf k}} b_{ {\bf k}} ), \]
where the spin wave excitation energy \( \epsilon({\bf k}) \) is:
\[	\epsilon ( {\bf k} ) = ( (g \beta H_{A} + 2 \mid J \mid Sz)^{2} -
4 J^{2} S^{2} \gamma^{2} ( {\bf k} ) )^{ \frac{1}{2} }  \]
and where:
\[	\gamma^{2} ( {\bf k} ) \equiv \sum_{<nn>} \exp (i {\bf k. \Delta }) \]
leads to two modes - degenerated modes. The anisotropy energy:
\[	\epsilon_{A} \equiv g \beta H_{A}   \]
and the exchange energy:
\[	\epsilon_{x} \equiv 2 J S z,   \]
give the gap energy:
\[	\epsilon (0) = ( \epsilon_{A} ( \epsilon_{A} + 2 \epsilon_{x} ) )^
{ \frac{1}{2} }.  \]

In the longwavelengths \( k \approx 0 \) with no anisotropy we obtain:
\[	\epsilon ( {\bf k} ) \approx (8z)^{ \frac{1}{2} } JSka.  \]
In the longwavelengths \( k \approx 0 \) with the anisotropy we obtain:
\[	\epsilon ( {\bf k} ) \approx ( \epsilon_{A}^{2} + 2 \epsilon_{A} \epsilon_{x} + 8 J^{2} S^{2}z k^{2} a^{2} )^{ \frac{1}{2} }.  \]
We can obtain the commensurate structure and the incommensurate structure via 	nonlinearities.
The spin fluctuations - correlations are:
\[	[ < \{ \delta S^{z}( \tau ) \delta S^{z}(0) \} > ],  \]
\[	[ < \{ \delta S^{+}( \tau ) \delta S^{-}(0) \} > ] . \]
The phase transitions depends on the phase boundaries. In antiferromagnetic regions in the amorphous matrix these boundaries may substantially depart behavior of spin waves from those which are obtained in very large (physicaly) infinite regions as concerning its characteristic length. Using the mean field we can describe behaviour of spin waves including the critical behavior. This (later) behavior was studied for
itinerant electrons in the antiferromgnet for Cr by \cite{FM}.

\section{RELAXATION AND LINEWIDTH.}

Relaxation and linewidth here are studied following \cite{Kra}
Local fields for time dependent perturbation are given by:
\[	{\it H'} = A {\bf I}. \delta {\bf S},   \]
\[	\delta {\bf S} \equiv {\bf S}(t) - <{\bf S}>.  \] 
The	NMR line profile \cite{KT}  is:
\[	I( \omega ) = \int_{- \infty }^{ + \infty} \exp (i \omega t -
\Psi (t) ) dt , \]
Let us note that this is a theoretical lineshape:
\[ \Psi(t) = ( \frac{A}{ \hbar } )^{2} \int_{ 0 }^{ t} (t - \tau )
( < \{ \delta S^{z}( \tau ) \delta S^{z}(0) \} > + \frac{1}{2} . \exp (
-i \omega_{0} \tau ) .< \{ \delta S^{+}( \tau ) 
\delta S^{-}(0) \} > ) d \tau .\]
The spin-relaxation rate is:
\[	\frac{1}{T_{1}} = \frac{1}{2} ( \frac{A}{ \hbar } )^{2} \int_{- \infty
 }^{ + \infty} \cos ( \omega_{0} t ) .< \{ \delta S^{+}( \tau ) 
\delta S^{-}(0) \} > dt. \]
In the limit of short electron spin time correlations \( \omega_{0} \tau_{e} 
\ll 1 \) we obtain for lineshape:
\[ \Psi(t) = ( \frac{A}{ \hbar } )^{2} \mid t \mid \int_{ 0 }^{+ \infty }
( < \{ \delta S^{z}( \tau ) \delta S^{z}(0) \} > + \frac{1}{2} . 
< \{ \delta S^{+}( \tau ) \delta S^{-}(0) \} > ) d \tau . \]
It is the Lorentzian lineshape with the halfwidth 
\( \Delta \omega_{\frac{1}{2} } \) =
\[	= \frac{1}{T_{2}} = \frac{1}{T_{2}^{,}} + \frac{1}{T_{1}^{,}}, \]
where:
\[	\frac{1}{T_{2}^{,}} = ( \frac{A}{ \hbar } )^{2} \int_{ 0 }^{+ \infty }
( < \{ \delta S^{z}( \tau ) \delta S^{z}(0) \} > ) d \tau , \]

\[	\frac{1}{T_{1}^{,}} = ( \frac{A}{ \hbar } )^{2} \int_{ 0 }^{+ \infty }
 \frac{1}{2} . ( < \{ \delta S^{+}( \tau ) \delta S^{-}(0) \} > ) d \tau . \] 

The spin-relaxation time is given as:
\[	\frac{1}{T_{1}} = \frac{1}{2} ( \frac{A}{ \hbar } )^{2} \int_{ 0
 }^{ + \infty} . < \{ \delta S^{+}( \tau ) \delta S^{-}(0) \} > d \tau , \]
\[  \frac{1}{T_{1}} = 2 ( \frac{1}{T_{1}^{,}} ) . \]
For \( T \gg T_{N} \), the paramagnetic region, 
the local field spectra \( \sim \) Gaussian distribution centered about
zero frequency:
\[ < \{ \delta S^{i}(t) \delta S^{i}(0) \} > = \frac{S(S+1)}{3} \exp (
- \frac{1}{2} \omega_{e}^{2} t^{2} ),  \]
here:
\[	\omega^{2}_{e} \equiv ( \frac{J}{ \hbar } )^{2} zS(S+1)  \]
and:
\[  \frac{1}{T_{1}} = \frac{1}{T_{2}} = ( 2 \pi )^{ \frac{1}{2} }
( \frac{ A}{ \hbar} )^{2} \frac{S(S+1)}{3} \frac{1}{ \omega_{e} } .  \]
Note that the anisotropy of A tensor leads to \( \frac{1}{T_{2}} \ne
\frac{1}{T_{1}} \) in general.
Also note that  without exchange we obtain:
\[  \frac{1}{T_{2}} \approx ( \frac{ A S}{ \hbar} ). \]

Relaxation time \( T \ll T_{N} \) and low temperature region 
at those temperatures T at which the spin wave  description is adequate:
there is strong angular dependence of \( T_{1} \) but not \( T_{2} \)
due to preferential direction of ordered spins. We find
\( T^{3} \) dependence of both \( T_{1,2} \) for \(  T_{AE} \ll
T \ll T_{N} \), and exponential decrease of both \( T_{1,2} \) for \(  T_{AE} \geq T \).

Types of processes contributing to nuclear relaxation are direct: \( E_{SW} = E_{Zeeman-nuclear} \), however \( \hbar \omega_{0} \approx 0.01K \rightarrow \) is usually negligible, and Raman: \( \mid \omega_{k} - \omega_{k'} \mid = \omega_{0} \).
Let us discuss the Raman scattering:
The Hamiltonian:
\[ {\it H} = A \sin ( \theta ) (I_{i}^{+} + I_{i}^{-} ) \sum_{ {\bf k,k'}}
\exp ( i {\bf (k-k').r_{i}} ) (u_{ {\bf k}}u_{ {\bf k'}} \alpha_{k}^{*} 
\alpha_{k'} + v_{ {\bf k}} v_{ {\bf k'}} \beta_{k} \beta_{k'}^{*}   ) \]
in small-k limit for \( \omega_{k} \) gives:
\[	\frac{1}{T_{1}} = \sin^{2}( \theta ) ( const ) ( \frac{T}{ T_{N}} )^{3}
\int_{ \frac{T_{AE}}{T}}^{ + \infty } \frac{ x dx }{ \exp(x) -1},\]
here \( \theta \) is an angle between the direction of spin antiferromagnetic 
alignment and the direction of nuclear quantization:
\[	const \equiv \frac{(A \Omega )^{2} \eta^{4} (S+1)^{4}}{ 81 \pi^{3}
\hbar b^{3} k T_{N} }. \]
Here \( \Omega \) is an atomic volume, 
\[ \eta \equiv \frac{3 k T_{N}}{2JzS(S+1)} \approx 1, \]
\[	b \equiv \frac{r^{2}_{nn}}{3}.... s.c. .\]
Note that for \( \theta = 0 \) the second order is needed \cite{Mit} to take into account: 
\[	\frac{1}{T_{1}} \sim ( const'' ) ( \frac{T_{A}}{ T_{N}^{4}} ), \]
as the Fig.1. in \cite{Kra} van Kranendonk shows T-dependence.

For \( T \gg T_{N} \) we obtain:
\[ \frac{1}{T_{1}} = \sin^{2}( \theta ) ( const ) ( \frac{T}{ T_{N}} )^{3} \]
and for \( T \ll T_{N} \):
\[ \frac{1}{T_{1}} = \sin^{2}( \theta ) ( const' ) ( \frac{T}{ T_{N}} )^{2}
. \exp ( - \frac{T_{AE}}{T} ), \]
where 
( const' ) \( \equiv \) ( const ). \( ( \frac{ \pi^{2}}{6} ) \).

In small-k limit for \( \omega_{k} \), spectral density of the 
fluctuating field extends to \( \omega \gg \omega_{0} \)  and we obtain:
\[	\frac{1}{T_{2}} = ( \frac{1+ \cos^{2}( \theta )}{2} )
( const ) ( \frac{T}{ T_{N}} )^{3}
\int_{ \frac{T_{AE}}{T}}^{ + \infty } \frac{ x dx }{ \exp(x) -1}. \]
Note here slight angle dependence only.

For \( T = T_{N} \) transition region \( \mid T - T_{N} \mid \ll T_{N} \)
the wave-dependent susceptibility \( \chi ( {\bf k} ) \) for
\( {\bf k} + {\bf Q} \approx {\bf Q} \equiv \frac{ \pi }{a} (1,1,1) \) in s.c. is given by:
\[ < \{ \delta S_{ {\bf k}}^{i}(t) \delta S_{ {\bf -k}}^{i}(0) \} > = \frac{kT}
{g \beta^{2}} \chi^{i}(k)exp( - \frac{t}{ \tau_{k}}), \]
where \( \tau_{k} \) is the characteristic decay time in the electronic
spin system, \( i= x,y,z \).

Let us discuss several examples.

For a cubic crystal with no magnetic field and no anisotropy:
\[ \frac{1}{T_{1}} = \frac{1}{T_{2}} = const. ( 2 \pi )^{ \frac{1}{2} }
( \frac{ A}{ \hbar} )^{2} \frac{S(S+1)}{3} \frac{1}{ \omega_{e} } .
( \frac{T_{N}}{T - T_{N}} )^{ \frac{1}{2} }, \]
\[	C \approx 10^{-1} , \]
which is valid for:
\[ \frac{ \omega_{0} }{ \omega_{e} } < \frac{T - T_{N} }{T_{N}} < 10^{-2} . \]

For cubic crystal in the magnetic field:
\[ H > 0 ... \rightarrow \chi_{\perp}(K_{0}), \chi_{ \parallel }(K_{0}) 
\rightarrow T_{ \perp}, T_{ \parallel } , \]
\[	T_{N} - T_{ \parallel } = 3( T_{N} - T_{ \perp } ), \]
\[	\frac{T_{N} - T_{ \parallel }}{T_{N}} \approx ( \frac{H}{H_{E}} )^{2}, \]
\[ \frac{1}{T_{1}} = const. ( 2 \pi )^{ \frac{1}{2} }
( \frac{ A}{ \hbar} )^{2} \frac{S(S+1)}{3} \frac{1}{ \omega_{e} } .
( \frac{T_{ \perp }}{T - T_{  \perp }} )^{ \frac{1}{2} }, \]
\[ \frac{1}{T_{2}} = \frac{const}{2}. ( 2 \pi )^{ \frac{1}{2} }
( \frac{ A}{ \hbar} )^{2} \frac{S(S+1)}{3} \frac{1}{ \omega_{e} } .(
( \frac{T_{ \parallel }}{T - T_{ \parallel}} )^{ \frac{1}{2} } +
( \frac{T_{ \perp }}{T - T_{ \perp }} )^{ \frac{1}{2} } ). \]

For tetragonal crystal with strong anisotropy ( \( H_{A} \gg H \) ):
\[ \frac{1}{T_{2, \parallel }} = \frac{const}{2}. ( 2 \pi )^{ \frac{1}{2} }
( \frac{ A}{ \hbar} )^{2} \frac{S(S+1)}{3} \frac{1}{ \omega_{e} } .(
( \frac{T_{N }}{T - T_{ N}} )^{ \frac{1}{2} } +
( \frac{T_{ \perp }}{T - T_{ \perp }} )^{ \frac{1}{2} } ), \]
\[ \frac{1}{T_{2, \perp }} = \frac{const}{4}. ( 2 \pi )^{ \frac{1}{2} }
( \frac{ A}{ \hbar} )^{2} \frac{S(S+1)}{3} \frac{1}{ \omega_{e} } .(
( \frac{T_{N }}{T - T_{ N}} )^{ \frac{1}{2} } + 3
( \frac{T_{ \perp }}{T - T_{ \perp }} )^{ \frac{1}{2} } ), \]
\[ \frac{1}{T_{1, \parallel }} = const. ( 2 \pi )^{ \frac{1}{2} }
( \frac{ A}{ \hbar} )^{2} \frac{S(S+1)}{3} \frac{1}{ \omega_{e} } .(
( \frac{T_{N }}{T - T_{ N}} )^{ \frac{1}{2} } ), \]
\[ \frac{1}{T_{1, \parallel }} = \frac{const}{2}. ( 2 \pi )^{ \frac{1}{2} }
( \frac{ A}{ \hbar} )^{2} \frac{S(S+1)}{3} \frac{1}{ \omega_{e} } .(
( \frac{T_{N }}{T - T_{ N}} )^{ \frac{1}{2} } +
( \frac{T_{ \perp }}{T - T_{ \perp }} )^{ \frac{1}{2} } ). \]

Let us discuss indirect nuclear spin-spin interaction.
It is important source of linewidth below \( T_{N} \).
A nucleus \( \rightarrow \) SW \( \rightarrow \) another nucleus,
There is no contribution to \( T_{1} \)and no transfer of energy from
	nuclear to the electronic spin system.
The Hamiltonian in this case is:
\[	{\it H}_{N} = - \frac{A^{2}S}{2} \sum_{j,R} F(R)(I_{j}^{+}
I_{j+R}^{-} + I_{j}^{-}I_{j+R}^{+} ), \]
here:
\[	F(R) \approx \frac{a}{R} \exp (- ( \frac{H_{A}}{H_{E}} )^{ 
\frac{1}{2}} \frac{R}{a} ), \]
\[	\frac{H_{A}}{H_{E}} \approx 0.1 - 0.01 , \]
\[	\Delta \omega \approx \Delta \omega_{paramagnet}( \frac{ \omega_{E}}{ \omega_{A} } )^{ \frac{1}{4}}.\]

\section{NMR and PARAMAGNETIC PHASE.}

In van Vleckov paramagnetic phase there are localized spins \cite{D} and \cite{A}, nature of the coupling should be known. An atom with a single electron outside closed shells forms such a localized spin, here spin-orbit coupling is negligible, the Hamiltonian for the magnetic interaction of the electron with the nucleus is:
\[	{\it H} = 2 \beta \gamma \hbar {\bf I}.( \frac{ {\bf l}}{r^{3}} -
\frac{ {\bf s}}{r^{3}} + \frac{ {\bf r(s.r)}}{r^{5}} + \frac{ 8 \pi {\bf s}
\delta ( {\bf r} )}{3}, \]
here \( \beta \) is the Bohr magneton.
Model for an atom with closed shell plus one electron with
\( \phi \) the orbital wave-function real \( \rightarrow :\)
\[ ( \phi \mid {\bf l} \mid \phi ) = 0 .\]
There is a tensor coupling:
\[ ( \phi \mid {\it H} \mid \phi ) = \hbar \gamma {\bf I}. {\it T}.{\bf S} ,\]
\( \phi = \sum_{l=s,p,d,...} a_{l} \phi_{l}  \)
s-part \( \rightarrow \) a scalar term A:
\[	= \hbar A {\bf I}.{\bf S}, \]
\[	A = \frac{16 \pi }{3} \beta \gamma \mid a_{0} \mid^{2} \mid \phi_{0}(0)
	 \mid^{2}. \]
Note that p,d, ..., -parts contribute:
\[	{\bf I}.{\it T'}.{bf S} = 2 \beta \sum_{l,l'} a_{l}a_{l'}^{*}
( \phi_{l'} \mid \frac{3}{r^{5}} {\bf (I.r)(S.r)} - \frac{1}{r^{3}} {\bf (I.S)}
\mid \phi_{l} ), \]
where:
\[	\mid l - l' \mid = 0, 2 . \]
Note also that:
if spin-orbit coupling is nonnegligible: Kramers theorem gives \( \rightarrow \)
still 2x degeneracy and \( \rightarrow \) fictious spin 1/2 .
Atoms with more than one electron outside closed shells have S, it is the total spin,
magnetic coupling of an electron with a nuclear spin that does not belong to the same atom is: 
\[	= \hbar A {\bf I}.{\bf S}  + 2 \beta \gamma \hbar {\bf I}.
grad_{R} \int \frac{ div( {\bf S} \rho (r) ) d^{3}r }{ \mid {\bf r - R } 
\mid }, \]
where \( \rho \equiv \mid \phi \mid^{2} :\)
\[	A = \frac{16 \pi }{3} \beta \gamma \mid a_{0} \mid^{2} \mid \phi_{0}(
 {\bf R} ) \mid^{2}. \]

\section{NMR and PARAMAGNETIC CONDUCTORS.}

Paramagnetic conducting phase is characterized by the coupling of
conduction electrons with the nuclear spins described by the same
Hamiltonian as in non-metals, but the conduction electrons are not localized, 
nuclear spin sees magnetic fields produced by all conduction electrons which form
a degenerate Fermi gas. If {\bf H} is an applied magnetic field and the gas with density of states on the Fermi level:
\[	g(E_{F}) = \frac{ 3NV}{2E_{F}}, \]
the magnetization is given by:
\[	M = \frac{ \beta n}{V} = \frac{\beta^{2} H g( E_{F} )}{V}, \]
and the paramagnetic susceptibility:
\[	\chi_{p} = \frac{M}{H} = \beta^{2} \frac{ g( E_{F} )}{V} =
\frac{ 3N \beta^{2} }{ 2 k T_{F}} .\]
The hyperfine coupling is:
\[	 = 2 \beta \gamma \hbar {bf I}. \sum_{unfilled orbits}
( \phi_{k} \mid (  - \frac{ {\bf s}_{k}}{r_{k}^{3}} + \frac{ {\bf r_{k}(s_{k}.
r_{k})}}{r_{k}^{5}} + \frac{ 8 \pi {\bf s_{k}}
\delta ( {\bf r}_{k} )}{3}, \]
\[	 = \gamma \hbar {bf I}. \sum_{unfilled orbits} {\it T}_{k}.{\bf s}_{k} \]
for \({\it T}_{k}  \approx  const \) near the top of the Fermi surface:
\[	 = \gamma \hbar {bf I}. {\it T}. \sum_{unfilled orbits}.{\bf s}_{k}. \]
Note that:
\[	2 \beta {\bf S} = -V {\bf M} = - V \chi_{p} {\bf H}_{0}, \]
thus:
\[	= -V \frac{ \gamma \hbar }{2 \beta } {\bf I}. \chi_{p} {\it T}.
{\bf H}_{0}, \]
which is the Knight shift:
\[	K = \frac{ \Delta H}{H_{0}} = \frac{8 \pi }{3} < \mid \psi_{k}(0) 
\mid^{2} >_{F} \chi{p,M}M . \]
Here K range from \( 2.5 x 10^{-4} ... Li^{7} \) to \( 2.5 x 10^{-2} ... 
Hg^{199} \),
all known values of K see in \cite{K} .

The Korringa law - numbers \cite{Mit} in the approximation of noninteracting conduction electrons the Knight shift  K and the spin lattice relaxation time 
\( T_{1} \) are related via Korringa relation  \cite{Sl} and \cite{Ca}:
\[	K^{2}TT_{1}= \frac{ \hbar}{ 4 \pi k_{B} } ( \frac{ \gamma_{e}}{\gamma_{n} } )^{2} ,  \]
where \( \gamma_{e}, \gamma_{n} \) are the electronic and nuclear gyromagnetic
ratios.

The electron-electron interactions within the conduction band modify the Korringa relation to:
\[	K^{2}TT_{1}= \frac{ \hbar}{ 4 \pi k_{B} } ( \frac{ \gamma_{e}}{ 
\gamma_{n} } )^{2} . \frac{1}{ K( \alpha ) }, \]
where \( K( \alpha \) is the enhancement factor smaller than 1,
see in \cite{Mor} - \cite{Mor2}.
Let us introduce some numbers for: ( \( 1 \equiv LaAl_{2}, 2 \equiv CeAl_{3} \) are:
for 1:
\[	T_{1}T=14 ( \pm 1 ) secK \rightarrow \frac{1}{K( \alpha ) } =1.34 , \]
for 2: see fig 10 \cite{LMac}. More details are described elsewhere on e-e interaction enhancement and on Kondo effects ( short range spin-spin correlations in the paramagnetic phase \cite{Sil} at sufficiently high temperatures such that critical fluctuation phenomena
are unimportant for normal metals.

\section{DISCUSSION.}

Behavior of the $Cu-Mn-Al$ ribbons was explained by the s-d interaction between conduction electrons and the clustered Mn atoms. While nuclear magnetic resonance measurements shows the antiferromagnetic and ferromagnetic clusters of Mn atom coexisting without long-range order, we described theory of NMR for AFM phase to study magnetic resonance properties also for the antiferromagnetic crystal phase regions (which have long-range order for larger regions) and which occur in these ribbons. The Heusler  Type  $Cu-Mn-Al$  Alloy has  a  composition  between $Cu_{2}MnAl$ and $Cu_{3}Al$. Electron  microscopy  of  the  premartensitic  $β-Cu-Zn-Al$  alloy has shown that the  $β-Cu-Zn-Al$  alloy  quenched  from  high  temperature  has the  electron  diffraction  patterns  of  this  alloy  well  explained  by  the model  with   the  existence of  small  particles  with  an  orthorhombic  structure.  It was noted that an important aspect of improvement in the material properties is to create a nanostructured state in matrix, which has significant advantages in magnetic and mechanical characteristics in contrast to the bulk materials in crystalline or amorphous state. Thus it is an interesting problem to study magnetic resonance properties not only for the Mn atoms and clusters without long-range order but also for the antiferromagnetic crystal phase regions (which have long-range order for larger regions in microcrystals and larger crystals) which may also occur in ribbons, besides nanocrystal regions. To study nanocrystal regions it is necessary to take into account more complicated magnetic structure of these nanocrystals.

Let us note that NMR tables and basic physical constants, as well as
conversion of Gaussian to SI Units are in \cite{Sl}, index of nuclear species is in \cite{A}. Tabular data of NMR in rare-earth intermetallic compouns is in \cite{D} and a survey of applications of NMR and NGR methods in \cite{Ba}.

Domain walls were studied in \cite{D}, they may occure in crystal regions of ribbons. In this case the zero-field NMR spectra of ferromagnetically ordered compounds are complicated due to presence of domain walls:
\[	H_{1,eff.} = H_{1} (1 + \eta ), \]
where enhancement factor for nuclei situated in Bloch walls and  domains
is \( \approx \) 100 - 10 000 but differs depending on the position, there exists broadening. Similar problem may occure in antiferromagnetic phase regions.

Let us note that several other problems are studied for NMR and AFM state.
The influence of the crystal electric field (CFE) on NMR spectra in AFM state are described in \cite{MC}. Here effect of spin fluctuations on the relaxation of a crystal-field-split rare-earth impurity is studied.
In \cite{D} the CFE influence on NMR spectra is studied experimentally
and theoretically.

While $Cu-Mn-Al$ materials are not ferromagnetic let us note that the NMR spectra for the ferromagnetic phase are studied and described in
\cite{Mit} and \cite{D}. There is  described the	ferromagnetic phase
hamiltonians for nuclei and for electrons. The spin waves are studied with 
spin-relaxation rate experimentally and theoretically.

On NMR and conduction electron polarisation has influence  which is studied, together with the Kondo phenomenon, in \cite{J1}, \cite{J2} and in
\cite{D}. Usually the uniform polarization model is used.
The RKKY-type analysis of the conduction electron polarization leads to description of correlation of magnetic ordering temperatures and
transferred hyperfine fields, the distance dependence of the transferred hyperfine interaction, the anisotropy of the transferred magnetic hyperfine interaction wand evidence for magnetically induced nuclear quadrupole interaction.

In some cases the AFM structure is incommensurate, see above. Such a case was studied for example in \cite{Det} for NMR proton line shape in \( (TMTSF)_{2}X \) where there is an incommensurability of nesting vector and order parameter, in \cite{GHHSHO} and \cite{HGO} for $CeAl_{2}$. The
lineshape in this case is described in \cite{Det} 
Analysis of the lines in the metallic state leads to:
\[	g( \omega , \Delta ) = \frac{1}{ (2 \pi )^{ \frac{1}{2} } \Delta}
\exp ( - \frac{ \omega^{2}}{ 2 \Delta^{2}} ). \]
Thus besides SDW in commensurate case also incommensurate case may be determined from the spectra.

In AFM and phonons and structures were studied 
experimentally and theoretically, and also 
AFM and phase transitions, critical phenomena (where a critical index is seen by NMR ),experimental and theoretical situations are described in \cite{Mor} - \cite{Mor2}.

In \cite{LCHBP} authors study noise characteristics of microwire magnetometer. Current trends lead to replacement of amorphous ribbon cores with magnetic microwires. However the miniaturization causes degradation in the parameters of sensors, so, considering measurement of weak magnetic felds, it is necessary to explore noise parameters, temperature drift and stability of the magnetometer output value. The article deals with analysis of microwire sensor noise characteristics based on the experimental data processing. In these magnetic wires are, however, metallic ribbons used on as the surface layers of wires to minimize degradtion. Then it is necessary again to study amorphous metallic ribbons, see also \cite{LCHBP}. As it is known an amorphous metallic surface layers improve these characteristics. To obtain optimal influence of the surface amorphous matallic layer (ribbon-like) it is necessary to understand its basic properties.

\section*{Acknowledgement.}
Some problems considered here were discussed by one of authors (O.H.) during his visit in the ETH Zuerich. He expresses his sincere thanks for discussions and for kind hospitality to Prof. H.R. Ott and to Dr. J. Gavilano. He acknowledges the financial support by the Swiss goverment grant.

\end{document}